\documentclass[prepint]{JHEP3} % 10pt is ignored!

\JHEPspecialurl{http://jhep.sissa.it/JOURNAL/JHEP3.tar.gz}

\usepackage{epsfig,multicol,bbm}

\newcommand\fverb{\setbox\fverbbox=\hbox\bgroup\verb}
\newcommand\fverbdo{\egroup\medskip\noindent%
			\fbox{\unhbox\fverbbox}\ }
\newcommand\fverbit{\egroup\item[\fbox{\unhbox\fverbbox}]}
\newbox\fverbbox

\def\laq{~\raise 0.4ex\hbox{$<$}\kern -0.8em\lower 0.62
ex\hbox{$\sim$}~}

\def\gaq{~\raise 0.4ex\hbox{$>$}\kern -0.7em\lower 0.62
ex\hbox{$\sim$}~}

\def\beq{\begin{equation}}
\def\eeq{\end{equation}}
\def\bea{\begin{eqnarray}}
\def\eea{\end{eqnarray}}

\def \pa {\partial}

\def \ra {\rightarrow}
\def \la {\lambda}
\def \La {\Lambda}

\def \da {\delta}
\def \ep {\epsilon}
\def \r {\rho}
\def \om {\omega}

\def \oh {{\overline h}_\chi}
\def \Ms {M_{\rm SUSY}}
\def \Mbr {M_{\rm S}^{\rm brane}}
\def \Mbu {M_{\rm S}^{\rm bulk}}
\def \Mp {M_{\rm P}}

\def \d {{\rm d}}
\def \e {{\rm e}}
\def \i {{\rm i}}

\title{Higher-dimensional perturbations of the vacuum energy density}

\author{M. Gasperini \\
	Dipartimento di Fisica, Universit\`a di Bari, 
Via G. Amendola 173, 70126 Bari, Italy,\\
and Istituto Nazionale di Fisica Nucleare, Sezione di Bari, 
Bari, Italy\\
	E-mail: \email{gasperini@ba.infn.it}}

\received{xx.xx.xx} 		%%
\accepted{xx.xx.xx}		%% These are for published papers.

\preprint{\hepth{yymmnnn}}	% OR:
\preprint{BA-TH/579-08\\
arXiv:yymm.nnnn\\
To appear in {\bf JHEP}}
\abstract{The vacuum energy density arising from the broken supersymmetry of the (standard-model) fields living on a brane cannot be fully ``off-loaded" to the bulk: even assuming the existence of an effective ``self-tuning" mechanism, a small fraction of the transferred energy ``bunces back" to the brane, as a backreaction of the supersymmetry breaking gravitationally transmitted to the bulk. In that case the SUSY scale of the brane has to be bounded, to guarantee the consistency of such a residual energy density with current large-scale phenomonological constraints. This effect is illustrated by computing the zero-point energies of the tower of (higher-dimensional) massive states associated to tensor metric fluctuations on a brane embedded in a warped bulk geometry, and it is shown to be independent of the number of compact or non-compact extra dimensions.}


\begin{document} 

%\maketitle  IS IGNORED %%%%%%%%%%%

\section{Introduction}

It is widely believed that exact supersymmetry is the most robust (presently known) mechanism able to protect a theory of fundamental interactions against the huge contribution of the quantum zero-point energy densities (thanks to the automatic Bose-Fermi cancellations), and to guarantee a vanishing cosmological constant. In principle, supersymmetry also allows stable vacua with a negative energy density $V<0$, but all known attempts to embed the supersymmetric standard model into string theory naturally lead to $V=0$ (see e.g. \cite{1}). 
However, supersymmetry appears to be broken in our real world, where the absence of observed superpartners for known particles suggests a breaking scale $\Ms \gaq 100$ GeV; even if such a breaking is spontaneous -- and the supertrace of the squared mass matrix of all supermultiplets is vanishing -- there is still a resulting failure of Bose-Fermi cancellations (see e.g. \cite{2}) which is then expected to produce a vacuum energy density $\r_V$ of the order of $\Ms^4 \gaq 10^{-64} \Mp^4$. This is striking contrast with the value suggested by current large-scale observations, $\r_V \laq 10^{-120} \Mp^4$ (see e.g. \cite{3}), where $\Mp= (8 \pi G)^{-1/2} \sim 10^{18}$ GeV is the (reduced) gravitational mass scale associated to the four-dimensional Newton constant.

A possible attractive way of reconciling a high SUSY breaking scale with a low energy density of the vacuum is in principle offered by the so-called brane world scenario, where -- according to the suggestions of open string theory \cite{4} and heterotic M-theory \cite{5} -- the non-gravitational interactions of our four-dimensional Universe are confined on the hypersurface swept by the time evolution of a three-brane embedded in a higher-dimensional ``bulk" manifold. In that case, as first suggested in \cite{6}, there are indeed ``warped" background solutions where -- thanks to fine-tuning \cite{7} or self-tuning \cite{8,8a} mechanisms -- the vacuum energy density of the fields living on the brane (and generating the intrinsic brane tension) is fully (or at least partially) absorbed by the bulk curvature along the spatial directions orthogonal to the brane (see also \cite{8b}).  In that context the brane geometry may even remain flat, or with a small enough cosmological constant \cite{2,8a,9,9a}, in spite of a realistic (i.e., high enough) supersymmetry breaking scale $\Mbr$ characterizing the standard-model fields confined on the brane. 

It is known that the above class of ``tuned" warped solutions has difficulties already at the level of the classical gravitational theory (see e.g. \cite{11}). Assuming that such difficulties are due to the effective low-energy approach, and may disappear at the level of an exact string theory description, we must consider however the stability of the brane geometry against the backreaction of bulk supersymmetry breaking, possibly generated by the brane itself (see e.g. \cite{12a,12}), and contributing back to the energy density of the brane through the zero-point energies of the fields living in the bulk. Indeed, for the consistency of the background solution, such a residual contribution has to be smaller than the primary brane energy density $\sim (\Mbr)^4$ absorbed by the bulk geometry. Also, and most important, the residual contribution should be sufficiently suppressed to be compatible with the observed value of $\r_V$, to avoid re-introducing the cosmological constant problem. 

The aim of this paper is to discuss this last point by computing the zero-point energies of the tower of massive states (of  extra-dimensional origin) associated to tensor metric fluctuations on  a brane embedded in a warped bulk geometry. Assuming that the bulk supersymmetry scale -- as well as the bulk curvature along the directions orthogonal to the brane -- are gravitationally  controlled by 
$\Mbr$ (as in models where bulk SUSY breaking directly originates  from the brane  \cite{8a,12a,12}), it is found that the vacuum energy density transferred from the bulk to the brane via gravitational backreaction is small enough to be compatible with a perturbative description. However, the phenomenological upper bounds on such a residual vacuum energy imply that the SUSY scale of the brane cannot be arbitrarily high: it turns out, in particular, that $\Mbr$ should not greatly exceed the TeV scale, quite independently of the number of compact or non-compact (warped) extra dimensions. This suggests that the idea of ``bulk-diluted" cosmological constant could be the object of observational tests very soon in forthcoming accelerator experiments (see e.g. \cite{13a}). 

\section{Zero-point energies of bulk gravitational fluctuations}

We will consider a very simple model of gravity described by the following higher-dimensional action, 
\beq
S= -{M_D^{D-2}\over 2} \int \d^D x \sqrt{-g}\, R, 
\label{1}
\eeq
where $M_D$ is the gravitational mass scale of the $D$-dimensional bulk manifold. We will assume that this action is complemented by the appropriate bulk and brane sources (see e.g. \cite{13}), so as to admit  $Z_2$-even background solutions describing a (possibly curved) $(D-1)$-dimensional hypersurface embedded into a ``warped" bulk geometry with metric
\bea &&
ds^2 =g_{AB} dx^A dx^B \equiv 
 f^2 (|z|) \left[ g_{\mu\nu}(x) dx^\mu dx^\nu - dz^2\right], \nonumber \\ 
 &&
 g_{\mu\nu}= {\rm diag} \left(1, - a^2 (t) \da_{ij}\right)
\label{2}
\eea
(conventions: $A,B=0,1, \dots, D-1$, and $\mu,\nu=0,1, \dots, D-2=d$). 
Here $x^A= (x^\mu, z)$, where $x^\mu=(t, x^i)$, with $i=1, \dots , d$, are the coordinates spanning the world-volume of the  $d$-brane (in the cosmic-time gauge), while $z$ is the coordinate along the orthogonal direction.

The metric $g_{\mu\nu}(x)$ provides a local parametrization of the intrinsic geometry of the brane world-volume, taking into account the possible sources of curvature left on the brane after the ``off-loading" of its vacuum energy density (or of part of it) into the external bulk geometry. The warp factor  $f(|z|)$ describes instead the ``bending" of the space-like dimension orthogonal to the brane, and is assumed to have a suitable $z$-dependence to guarantee the convergence of the integral
\beq
\int_{-\infty}^{+\infty} \d z f(|z|)^{D-2}=  L,
\label{3}
\eeq
assigning a typical ``proper size" $L$ to this extra dimension (and providing a finite relationship between the bulk and the brane gravitational mass scales, $M_D^{D-2}  L \sim \Mp^{D-3}$). In particular, the metric (\ref{2}) could describe a globally flat (Minkowski) hyperplane embedded into an AdS bulk manifold (as in the example considered in \cite{7}); however, the explicit forms of $a(t)$ and $f(|z|)$ are irrelevant for the purpose of this paper. Note that we could also include in our discussion the case in which the dimension external to the brane is flat  (i.e. $f=$ const), provided it is compact (as in the conventional Kaluza-Klein scenario) and of size $L$. 

Let us now perturb the background geometry (\ref{2}),  $g_{AB} \ra g_{AB} + \da g_{AB}$,  focusing our attention on the transverse  and traceless part of the metric fluctuations along directions parallel to the world volume of the brane, i.e. considering the perturbed configuration characterized by 
\beq
\da g_{AB}= h_{AB}, ~~~~~ h_{Az}=0, ~~~~~ h_{\mu\nu}= h_{\mu\nu}(x^\mu, z) \not=0, ~~~~~ g^{\mu\nu}h_{\mu\nu}=0= \nabla^\nu h_{\mu\nu}
\label{4}
\eeq 
(in the linear approximation, the various components of $h_{AB}$ are decoupled from each other, and can be treated independently). Expanding the action (\ref{1}) around the background (\ref{2}) up to terms quadratic in $h_{\mu\nu}$, and using the synchronous gauge 
$h_{0\mu}=0$, $h_{ij}\not=0$, we then obtain the following quadratic action (see e.g \cite{14}), 
\beq
S^{(2)}= -{M_D^{D-2}\over 8}  \int \d^D x \sqrt{-g}~ h_i^j \nabla_A\nabla^A h_j^i , 
\label{5}
\eeq
where $\nabla_A\nabla^A$ is the covariant d'Alembert operator in $D$ dimensions. Integrating by parts, and setting   $h_{ij}= h_a\ep_{ij}^a$, where $\ep_{ij}^a$ is the spin-two polarization tensor satisfying the trace property Tr\,$(\ep^a \ep^b)=2 \da^{ab}$, we are lead to 
\beq
S^{(2)}= \sum_a S_a, ~~~~~~ S_a=
{M_D^{D-2}\over 4}  \int \d z\int \d^{d+1}x \,a^d f^d \left[\dot h_a^2-\left(\pa_i h_a\over a\right)^2 -h_a^{\prime2}\right],
\label{6}
\eeq
where the dot denotes a time derivative,  the prime a derivative with respect to the extra coordinate $z$, and the sum is over all the  independent polarization states. Considering an unpolarized fluctuation background we will concentrate our subsequent discussion on a single polarization mode $h$ (omitting, for simplicity, the polarization index). By varying  the action (\ref{6}) with respect to $h$ we can finally obtain the linear evolution equation for the components of tensor metric perturbations parallel to the brane, 
\beq
\ddot h+d {\dot a \over a}\dot h- {\nabla^2\over a^2} h-h''-d {f'\over f} h'=0.
\label{7}
\eeq
where $\nabla^2= \da^{ij} \pa_i \pa_j$. 

The above equation, together with the action (\ref{6}), provides  the required starting point for the canonical normalization of bulk gravitational fluctuations, and for the computation of their contributions to the vacuum energy density of the brane. To this purpose, let us first  separate the coordinate dependence in Eq. (\ref{7}) by setting $h(x^\mu,z)=v_\chi(x^\mu)\psi_\chi(z)$, thus obtaining the  eigenvalue equations:
\bea
&&
\psi_\chi''-d {f'\over f}\psi_\chi' \equiv f^{-d} \left(f^d \psi_\chi'\right)'=- \chi^2 \psi_\chi,
\label{8} \\
&&
\ddot v_\chi + d {\dot a \over a} \dot v_\chi - {\nabla^2\over a^2} v_\chi= -\chi^2 v_\chi.
\label{9}
\eea
Also, let us suppose that the massless mode, corresponding to  $\chi=0$ and to $\psi_0=$ const, is strictly localized on the brane (as in the example of \cite{7}), while the massive fluctuations are free to propagate in the bulk, and characterized by a continuous spectrum of values of $\chi^2$. For the massive modes we can then write the general solution of Eq. (\ref{7}) in the form 
\beq
 h(x^\mu,z)=L\int \d \chi\,v_\chi(x^\mu)\psi_\chi(z),
 \label{10}
 \eeq
where the factor $L$ has been introduced to keep $v$ and $\psi$ dimensionless, and where the ``eigenfunctions" $\psi_\chi$ of Eq. (\ref{8}) are normalized with respect to inner products with measure $dzf^d$, 
\beq
\int \d z\, f^d \psi_\chi\psi_{\chi'}= \da (\chi+\chi').
\label{11}
\eeq
This normalization is obtained by imposing on the rescaled variable $\widehat \psi_\chi= f^{d/2} \psi_\chi$ -- satisfying Eq. (\ref{8}) in canonical, ``Schrodinger-like" form -- to be normalized with measure 
$\d z$ as in conventional one-dimensional quantum mechanics (see e. g. \cite{7,15}). Note that in the absence of warping (i.e. for $f=$ const) the solutions of Eq. (\ref{8}) can be written in the plane wave form, $\psi_\chi =A_\chi \exp (\i \chi z)$, so that Eq. (\ref{10}) reduces to the standard definition of Fourier transform, and Eq. (\ref{11}) corresponds to the continuous $\da$-function normalization of plane waves in conventional quantum mechanics. For a compact extra dimension, however, the solutions satisfy periodic boundary conditions, the spectrum of $\chi^2$ is discrete, the integral $L \int \d \chi$ in Eq. (\ref{10}) is replaced by a sum over a dimensionless index, and the normalization of $\psi_\chi$ must be expressed in terms of the Kronecker symbol. 

Let us now insert the expansion (\ref{10}) into the action (\ref{6}). Using Eqs. (\ref{8}), integrating over $z$ with the help of the orthonormality condition (\ref{11}), and defining the field $\oh$
representing the amplitude of the massive bulk fluctuations evaluated at the brane position,
\beq 
\oh (x^\mu)= \left[h_\chi(x^\mu, z) \right]_{z=z_{\rm brane}}, 
\label{12}
\eeq 
we obtain (modulo a total derivative) the effective action (see e.g. \cite{14,16})
\bea
&&
S_a= L\int \d \chi S_\chi, 
\nonumber \\
&&
S_\chi= {M_D^{d}L\over 4 |\psi_\chi(z_{\rm brane})|^2}  \int\d^{d+1}x \,a^d\left(
\left|\dot \oh\right|^2- {\left|\nabla \oh\right|^2\over a^2}-\chi^2\left| \oh\right|^2\right).
\label{13}
\eea
Introducing the canonical variable $u_\chi$, such that 
\beq
u_\chi(x^\mu)= \xi\,\oh(x^\mu), ~~~~~~~~~~~~~~
\xi=\left(M_D^dL\over2\right)^{1/2}{a^{d/2}\over 
 |\psi_\chi(z_{\rm brane})|} ,
\label{14}
\eeq 
we finally arrive at the canonical action describing the contribution of the massive spectrum of bulk gravitational fluctuations, minimally (i.e. geodesically) coupled to the cosmological geometry of the brane:
\beq
S_\chi= {1\over 2} \int\d^{d+1}x  \left(
\left|\dot u_\chi\right|^2- {\left|\nabla u_\chi\right|^2\over a^2}-\chi^2\left| u_\chi\right|^2-V(\xi)\left| u_\chi\right|^2\right), 
\label{15}
\eeq
where 
\beq
V(\xi)= -{\d \over \d t} \left(\dot \xi\over \xi\right) -{\dot \xi^2 \over \xi^2}
= -{d\over 2} \dot H -{d^2 \over 4} H^2, ~~~~~~~~~~~
H={\dot a \over a}.
\label{16}
\eeq

We are interested in the zero-point energies of these massive modes, which are determined by the free oscillations of the Fourier components $u_\chi(t,p)$,
\beq 
\ddot u_\chi(p)+ \om_\chi^2 u_\chi(p)=0, ~~~~~~~~~~~~
\om_\chi^2=p^2+\chi^2, 
\label{17}
\eeq
obtained in the adiabatic limit where the coupling to the geometry becomes negligible, $\dot a \ra 0$, $ V(\xi) \ra 0$. Using the positive frequency solutions of Eq. (\ref{17}), $u_\chi(t,p)= {\e^{-\i \om_\chi t}/\sqrt{2 \om_\chi}}$ (canonically normalized so as to satisfy  $u(p) \dot u^\ast(p)- \dot u(p) u^\ast(p)=\i$, and representing the vacuum state for quantized fluctuations, see e.g. \cite{17}), the computation of the (averaged) $T_{00}$ component of the canonical stress tensor for the free field solution then gives us the zero-point contribution to the energy density of the $d$-brane, 
\beq
\r_\chi={1\over (2 \pi)^d} \int \d^d p\, {1\over 2} \left(p^2+\chi^2\right)^{1/2}. 
\label{18}
\eeq
Summing up all $\chi$-mode contributions (according to Eq. (\ref{13})) we finally obtain the total induced vacuum energy density  
\beq
\r_V= L\int \d \chi \, \r_\chi= {L\over (2 \pi)^d}\int_0^\la \d \chi \int_0^\La \d^d p\, {1\over 2}\left(p^2+\chi^2\right)^{1/2}, 
\label{19}
\eeq
where we have introduced the (possibly different) cutoff parameters $\La$ and $\la$ in the momentum spaces associated to the spatial dimensions internal and external to the brane, respectively. 

\section{Backreaction of bulk supersymmetry breaking}

From now on we shall assume that the $d$-brane represents our ordinary macroscopic world, so that $d=3$, and that the bulk supersymmetry is broken at a given scale $\Mbu$, so that the zero-point energies of the bulk gravitational fluctuations are not exactly cancelled by other fields present in the supersymmetric multiplet.  Subtracting from Eq. (\ref{19}) the associated contribution of the bulk fermionic partners, considering a model of broken supersymmetry where there is an equal number of boson and fermion degrees of freedom, and assuming the (supertrace) cancellations of the mass-squared terms \cite{2}, we can then obtain from Eq. (\ref{19}), to leading order, 
\beq
{\r_V \over \Mp^4} \sim L \Mbu \left( \Mbu \over \Mp\right)^4. 
\label{20}
\eeq
We may note that $ \la \sim \Mbu$, since above that scale the supersymmetry is restored, and appropriate cancellations are expected to suppress to zero the contribution of bulk fluctuations to the vacuum energy density. 

It is important to stress that the prefactor 
$ L \Mbu$ in the above equation is peculiar of a continuous fluctuation spectrum\footnote{I am indebted to G. Veneziano for a useful discussion on this point.} (and thus of a non-compact, warped extra dimension). In fact, if we have a discrete spectrum of momenta $\chi_n$ (associated to a compact extra-dimension of size $L$, sufficiently small so that the eigenvalue spacing $L^{-1}$ is not negligible with respect to the bulk scale $\Mbu$), then the integral $\int \d \chi$ of Eq. (\ref{19}) is replaced by the (dimensional) sum operator $L^{-1} \sum_n$, and the new result for $\r_V$ -- assuming the convergence of the series of SUSY breaking corrections at the bulk scale, as before -- is simply $\r_V \sim (\Mbu)^4$.

This result can also be understood by noting that, for compact dimensions of size $L^{-1} \gaq \Mbu$, the associated vacuum energy density is dominated by the Casimir effect, whose contribution is inversely proportional to the volume of the compact space. For a single compact dimension, in particular, the five-dimensional energy density of the bulk fluctuations becomes $\sim L^{-1} (\Mbu)^4$, and the overall contribution to the brane energy density (obtained by integrating over the proper volume of the transverse compact dimension) is  $\r_V \sim (\Mbu)^4$. In the limit of a (warped) non-compact dimension, on the contrary, the Casimir energy becomes negligible, the five-dimensional energy density is simply $(\Mbu)^5$ (irrespectively of the relative magnitude of $L^{-1}$ and $\Mbu$), and the integration over the transverse volume leads to the result of Eq. (\ref{20}). The above arguments can be extended to the case in which there is a number $n=D-4>1$ of space-like dimensions external to the brane, to obtain that the leading contribution $(\Mbu)^4$ is multiplied 
 by a factor $ L \Mbu$  {\em for any} of the non-compact extra dimensions present in the bulk and characterized by a continuous spectrum of fluctuations (we are assuming, for simplicity, that they are all of the same size $L$). Hence 
\beq
{\r_V \over \Mp^4} \sim\left(L\Mbu\right)^N \left( \Mbu \over \Mp\right)^4,
\label{21}
\eeq
where $N$ is the total number of non-compact, warped extra dimensions.  
Note that this number may also include the contribution of compact (but large enough) dimensions, whose mass-eigenvalue spacing $L^{-1}$ is negligible with respect to the energy scale $\Mbu$, so that the associated spectrum can be regarded as a continuous one. 

Up to now we have treated $L$ and $\Mbu$ as arbitrary parameters, useful for characterizing the intensity of the vacuum energy density  transmitted to the brane by the process of bulk supersymmetry breaking. At this stage, however, it is time to recall that we are  considering a ``tuned" background configuration where the typical curvature scale of the extra space-like dimensions is gravitationally controlled by the intrinsic energy density of the  brane (mainly due to the SUSY breaking of the fields confined on it), so that
\beq
L^{-2} \sim G \,\r_{\rm brane} \sim {(\Mbr)^4\over \Mp^2}
\label{22}
\eeq
(see e.g. \cite{7}). In the presence of bulk contributions, assuming that all $n$ extra dimensions have the same size $L$, and that there are $N$ dimensions ($0 \leq N\leq n$) with an energy density dominated by the bulk SUSY scale ($\Mbu \gaq L^{-1}$), while the remaining $n-N$ compact dimensions are dominated by the Casimir energy density 
($L^{-1} \gaq \Mbu$), we can also rewrite the previous equation as 
\beq
L^{-2} \sim G_{4+n} \, \r_{\rm bulk} \sim {(\Mbr)^4\over M_D^{2+n}}
{(\Mbu)^N\over L^{n-N}}.
\label{23}
\eeq
Here $G_{4+n} = M_D^{-2-n}$ is the bulk gravitational parameter, related to the Planckian gravitational coupling of the brane by
\beq
M_D^{2+n} L^n \sim \Mp^2. 
\label{24}
\eeq
We should also take into account that, even in the absence of specific sources of  SUSY breaking, a curvature of order $L^{-1}$ necessarily breaks  bulk supersymmetry at a scale $\Mbu$ of the same order (see e.g. \cite{8a,12a,12}). Assuming that this is the case, it follows that this minimal level of bulk SUSY breaking should be related to $\Mbr$, according to the above equations, as 
\beq
L \Mbu \sim 1, ~~~~~~~~~~~~~
{\Mbu\over \Mp} \sim 
\left(\Mbr\over \Mp\right)^2 .
\label{25}
\eeq
Inserting these conditions into Eq. (\ref{21}) we can finally conclude that the minimal, extra-dimensional vacuum energy density, absorbed by the brane as a gravitational backreaction of bulk SUSY breaking, is given by
\beq
{\r_V \over \Mp^4} \sim\left(\Mbu\over \Mp\right)^4 \sim \left( \Mbr \over \Mp\right)^8. 
\label{26}
\eeq
As anticipated, the result is  independent of both the number and the compactness of the external dimensions. 

\section{Conclusion}
The above estimate coincides with a result already suggested (with different arguments) in the literature \cite{2,8a,9,9a,18}. In our context, assuming the existence of some mechanism able to off-load to the bulk the leading-order energy density $(\Mbr)^4$, the residual energy (\ref{26}) determines the maximal allowed value of the parameter $\Mbr$ in a realistic brane-world scenario where the vacuum energy density  is bounded by current large-scale observations as 
$\r_V/\Mp^4 \laq 10^{-120}$.  Applying to Eq. (\ref{26}) such observational constraint we obtain the bound $\Mbr \laq 1$ TeV. This seems to  suggest  that the non-observation of supersymmetric effects in the planned, near-future collider experiments (where the TeV scale is available to experimental tests), should be interpreted as evidence  against the {\em naive} scenario of ``SUSY-breaking generated" and ``bulk-diluted" cosmological constant considered in this paper. 

It should be noted, as a final comment, that the bulk dimensionality does not affect the allowed value of $\Mbr$, but it is relevant to the value of the bulk gravitational scale $M_D$, which turns out to be related to $\Mbr$ by $M_D/\Mp \sim \left(\Mbr/\Mp\right)^{2n/(2+n)}$ (according to Eqs. (\ref{24}), (\ref{25})). Thus, $M_D \sim \Mbr$ only for $n=2$, as in the context of $6$-dimensional supergravity models discussed in \cite{8a}. It should be stressed, however, that the  coupling strength governing the interactions of the extra-dimensional massive bulk gravitons on the brane is defined by the effective action (\ref{13}): such a coupling depends not only on $M_D$, but also on the eigenfunction $\Psi_\chi(z)$ evaluated at the brane position, namely on a parameter which  is strongly model-dependent (see e.g. \cite{14}-\cite{16}).

\section{Acknowledgements}
It is a pleasure to thank  Gabriele Veneziano for many instructive discussions and many hints on the problems discussed in this paper.

\end{document}